\documentclass[traditabstract]{aa}

\usepackage{txfonts}
\usepackage{natbib}
\bibpunct{(}{)}{;}{a}{}{,} 

\usepackage[pdftex]{graphicx}
\usepackage{epstopdf}
\usepackage{longtable,lscape}

\newcommand{\linebreakcell}[2][c]{%
  \begin{tabular}[#1]{@{}c@{}}#2\end{tabular}}

\begin{document}
\title{A unified supernova catalogue\thanks{Full Table A.1 is only available at the CDS via anonymous ftp to         cdsarc.u-strasbg.fr (130.79.128.5) or via http://cdsarc.u-strasbg.fr/viz-bin/qcat?J/A+A/538/A120}}
\author{D.~Lennarz\inst{1}\thanks{\emph{Present address:} Max-Planck-Institut f\"ur Kernphysik, P.O. Box 103980, 69029 Heidelberg, Germany} \and D.~Altmann\inst{1} \and C.~Wiebusch\inst{1}}

\institute{III.Physikalisches Institut, RWTH Aachen University, 52056 Aachen, Germany}
\date{Received 8 July 2011 / Accepted 19 September 2011}

\abstract{In this paper a new supernova catalogue containing data for 5526 extragalactic supernovae that were discovered up to 2010 December 31 is presented. It combines several catalogues that are currently available online in a consistent and traceable way. During the comparison of the catalogues inconsistent entries were identified and resolved where possible. Remaining inconsistencies are marked transparently and can be easily identified. Thus it is possible to select a high-quality sample in a most simple way. Where available, redshift-based distance estimates to the supernovae were replaced by journal-refereed distances. Examples of statistical studies that are now possible with this new catalogue are presented in this paper.}

\maketitle
\titlerunning{A unified SN catalogue}
\authorrunning{Lennarz et al.}

\section{Introduction}
The observation and study of supernovae (SNe) have had significant scientific impact in the past 30 years. Most prominently, \object{SN1987A} enabled probing neutrino properties such as mass \citep{SNIa_neutrino_mass}, electric charge \citep{SNIa_neutrino_charge} and magnetic moment \citep{SNIa_neutrino_magnetic_moment}. Recently, SNe of type Ia have been used as ``standard candles'' to measure distances up to cosmological scales \citep{SNIa_distances_calibration}. In contemporary scientific research SNe are still important. Young SNe for instance may be identified as cosmic-ray accelerators by measuring high-energetic gamma or neutrino radiation shortly after the SN explosion \citep{pulsarmodel,andobeacom,rmwaxman}.

Zwicky started the first systematic SN search in 1932 at Caltech, but he and his collaborators detected the first SNe only after the newly built 18\arcsec Schmidt telescope at Palomar Observatory was put into operation \citep{zwicky_first_systematic_search}. A first overall SN list was published by \citet{zwicky_first_catalogue} and contained 54 SNe discovered between 1885 and 1956. This Palomar Supernova Master List has been updated and occasionally republished \citep{zwicky1965,kowal1971,sargent1974}. Two other SN lists were published contemporaneously \citep{karpwicz1968,flin1979}.

A new and revised supernova catalogue compiled from the Palomar Supernova Master List has been published by \citet{barbon1984}. This catalogue then became the Asiago SN catalogue \citep[ASC;][]{barbon1989,barbon1999} and has received great recognition so far. In 1993 another catalogue, the Sternberg Astronomical Institute (SAI) SN catalogue (SSC) was first published \citep{SSC1993,SSC2004}.

Today, running SN catalogues are easily accessible over the internet. The most important ones are the list of SNe maintained by the Central Bureau for Astronomical Telegrams (CBAT)\footnote{http://www.cfa.harvard.edu/iau/lists/Supernovae.html}, the electronic version of the ASC\footnote{http://web.oapd.inaf.it/supern/cat/} and the electronic version of the SSC\footnote{http://www.sai.msu.su/sn/sncat/}. The latter two can also be accessed via VizieR\footnote{http://vizier.u-strasbg.fr/cgi-bin/VizieR}.

\begin{figure}[t]
\includegraphics*[width=\columnwidth]{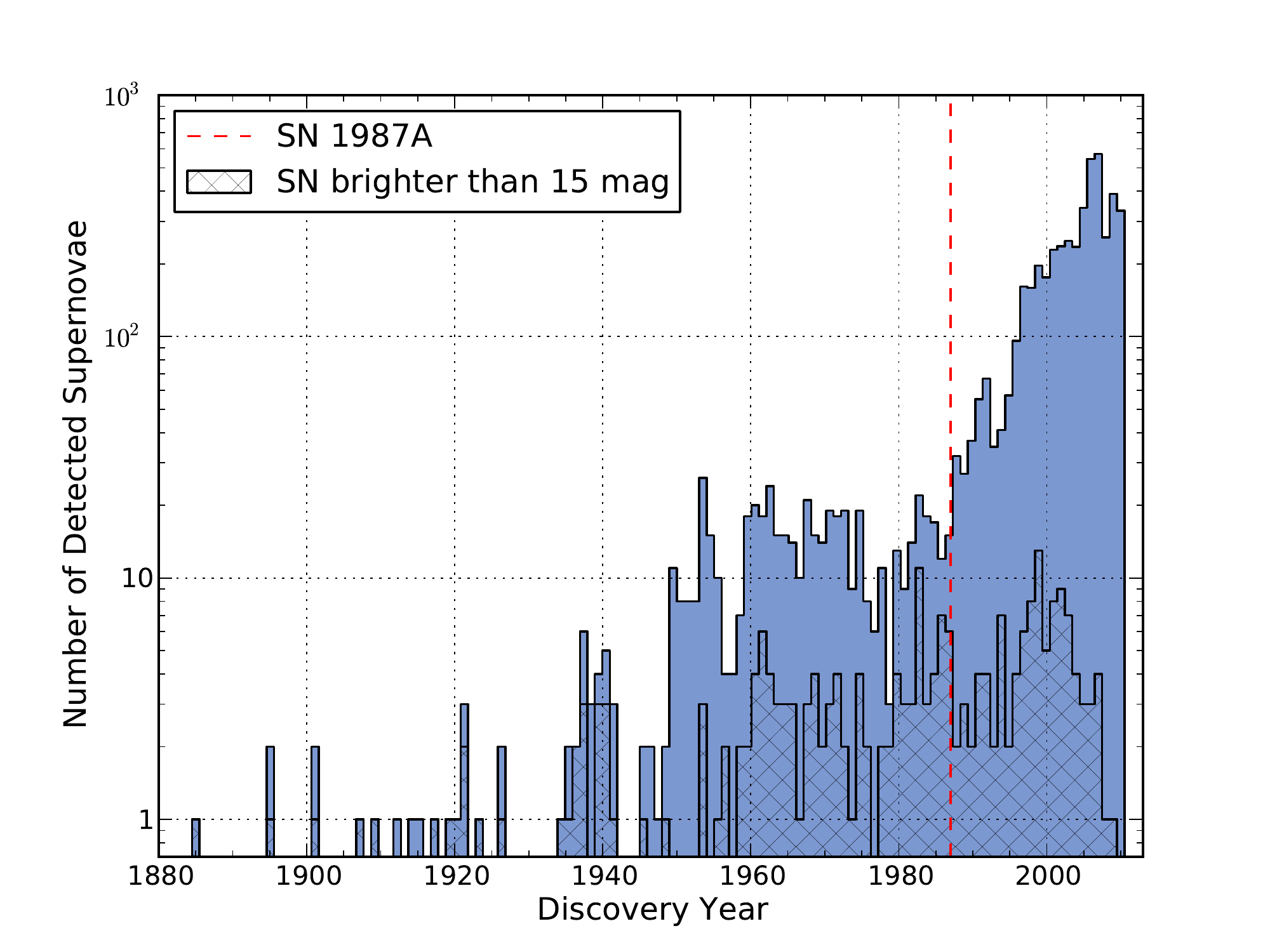}
\caption{Number of detected SNe per year. The discovery year of \object{SN1987A} is marked with a dashed line. The fraction of bright SNe, which have a magnitude at maximum $ < 15^{\rm{m}}$, is indicated by the hatched area. Selection criteria for the data are described in Sect. \ref{sec:studies}.}
\label{fig:snperyear}
\end{figure}

The recently growing interest in SNe is demonstrated in Fig.~\ref{fig:snperyear}. The number of detected SNe per year has grown almost exponentially after the discovery of \object{SN1987A}, reaching a current discovery rate of a few hundred SNe per year. This gain can be mostly attributed to distant SNe, because the discovery rate of bright SNe has not significantly increased.

The need for a new supernova catalogue has arisen during a search for high-energetic neutrinos from young SNe with the AMANDA neutrino telescope \citep{arxiv,Diplomarbeit}. In order to enhance the sensitivity of the analysis, the SNe were stacked (i.e. adding up small signals that may not be significant individually). Nearby SNe contribute most to the expected signal, but the distance derived from the redshift of the SN host galaxy gives only a very rough estimate. Hence, a new catalogue was compiled that includes journal-refereed distances of the host galaxies and therefore allows a more realistic signal estimate.

A second motivation for a new catalogue was that a comparison between the three SN catalogues (CBAT, ASC, SSC) available online revealed significant inconsistencies in the listed information. In some cases these can be attributed to simple typographical errors, while others are qualitative differences in the given information. The concept of the new catalogue is to list the undisputed information and flag differences. Therefore, the new catalogue also serves as a meta-catalogue of the current online SN catalogues. A subset of high-quality SNe with more reliable information can be selected with the meta-data.

\section{The unified supernova catalogue}
This chapter briefly describes the columns of the unified supernova catalogue (USC). In the next chapter a more detailed description of the data reduction and unification is given. An excerpt of the USC is shown in appendix \ref{tab:exam}, whereas the full catalogue is available at the address indicated earlier.

The SN data are taken from the three running SN catalogues that are available online (CBAT, ASC, SSC), last downloaded on 2011 June 1. All host-galaxy-related information is taken from HyperLeda\footnote{http://leda.univ-lyon1.fr/} (downloaded on 2011 June 1). The distances to the SN host galaxies (e.g. derived from Cepheid variables or the Tully-Fisher relation) were taken from the NASA/IPAC Master list of galaxy distances\footnote{http://nedwww.ipac.caltech.edu/Library/Distances/} (NED-D, version 4.1). It contains $36\,411$ distances to $9193$ galaxies.

Compared to the other catalogues, the USC contains less columns with data on the host galaxy. This decision was made to avoid data duplication, that is, the repetition of information from other sources that are only secondary to the SN data. Most galaxy information is accessible online and can be easily collected for the need of each analysis.

The columns in the USC are:
\begin{description}
  \item[\textbf{(1) Supernova designation}] The SN designation is followed by a varying number of asterisks indicating the uncertainty. An uncertainty can arise because the SN was not confirmed by other observers or because it was not a genuine SN (discussed in more detail in the next chapter). The meaning of the number of asterisks is:
      \begin{description}
        \item[*] SN flagged uncertain or not listed in the SSC,
        \item[**] SN flagged uncertain or not listed in the ASC,
        \item[***] SN flagged uncertain or neither listed in the ASC or SSC.
      \end{description}
  \item[\textbf{(2) Supernova host galaxy}] The SN host galaxy is crucial because the properties of the host galaxy are important for statistical studies and it can be used to obtain a redshift-independent distance estimate. Therefore, the host galaxies were carefully selected. The provided galaxy identification is the HyperLeda principal identifier if the host galaxy is known. A host galaxy in brackets means that this host galaxy is only given in the SSC. Presumably, the SSC uses the closest known galaxy to the SN position as host, but this information should be treated with caution (e.g. the real SN host galaxy may not have been catalogued). A host galaxy followed by an asterisk means that for this galaxy the parameter ``objtype'' in HyperLeda is different from ``G'' for galaxy or ``Q'' for a quasi-stellar object (QSOs). Currently, all these objects are marked as an ``extended source of unknown/uncertain nature''.
  \item[\textbf{(3) Supernova position}] The SN right ascension and declination is given in J2000.0 coordinates. The meaning of the number of asterisks is:
      \begin{description}
        \item[*] SN positions are inconsistent between catalogues by more than 5\arcsec,
        \item[**] SN position was obtained directly from a specific publication and was manually inserted into the catalogue
        \item[***] SN position was calculated from the host galaxy position and the SN offset (see next chapter).
      \end{description}
  \item[\textbf{(4) Supernova offset from host galaxy}] The offset gives the difference between the SN position to the core of the host galaxy in units of arcseconds. It follows astronomical conventions in the sense that ``W'' (west) means negative in the direction of the right ascension and ``E'' (east) means positive. For the declination ``N'' (north) means positive and  ``S'' (south) means negative direction. In order to convert the right ascension offset to a position, it has to be divided by the cosine of the declination and converted from arcseconds to seconds of time. If the SN position was added from a specific publication and the host galaxy position is known, the offset contains no additional information. In these cases it was replaced by ``irrel.'' (short for irrelevant). The meaning of the number of asterisks is:
      \begin{description}
        \item[*] SN offset was obtained directly from a specific publication and was manually inserted into the catalogue,
        \item[**] SN offset comes from one catalogue only,
        \item[***] SN offsets are inconsistent between catalogues by more than 5\arcsec.
      \end{description}
  \item[\textbf{(5-7) Supernova type}] The supernova type (Ia, Ib, Ic, II) is given in column 5. In cases where the information is inconsistent and unification could not be achieved, the different types from all catalogues are given separated by ``$|$''. The meaning of the number of asterisks is:
      \begin{description}
        \item[*] type is flagged uncertain in at least one catalogue,
        \item[**] type information comes from one catalogue only,
        \item[***] at least one catalogue does not list the subtype (e.g. one catalogue I, the other two Ia).
      \end{description}
      Column 6 lists cases where the light curve was peculiar with an ``x''. In cases where the peculiarity is uncertain, an asterisk was added. Column 7 lists the SN II subtypes (n, L, b, P). If this subtype is uncertain, the entry is also followed by an asterisk.
  \item[\textbf{(8) Discovery date}] If known, the discovery date of the SN is given. In cases where the ASC and SSC list the discovery magnitude for different photometric bands, both dates are given separated by ``$|$'' unless they are equal. The number of asterisks is additive:
      \begin{description}
        \item[*] value comes from less than three catalogues,
        \item[**] difference $\leq$10 days,
        \item[****] difference $>$10 days.
      \end{description}
  \item[\textbf{(9) Discovery magnitude}] If known, the (optical) discovery magnitude is given. In cases where the ASC and SSC list the discovery magnitude for different photometric bands, both magnitudes are given separated by ``$|$'' unless they are equal. The number of asterisks is additive:
      \begin{description}
        \item[*] value comes from less than three catalogues,
        \item[**] difference $\leq0.5^{\rm{m}}$,
        \item[****] difference $>0.5^{\rm{m}}$.
      \end{description}
  \item[\textbf{(10) Discovery magnitude band}] If known, the photometric band of the discovery magnitude is given. This value is only specified in the ASC and SSC. An asterisk after the magnitude band means that this value is taken from one catalogue only. If the ASC and SSC list different bands, both bands are given separated by ``$|$''.
  \item[\textbf{(11) Maximum date}] In some cases the ASC and SSC list the date when the light curve of the SN reached its maximum. In cases where the ASC and SSC list the maximum magnitude for different photometric bands, both dates are given separated by ``$|$'' unless they are equal. The number of asterisks means:
      \begin{description}
        \item[*] value comes from one catalogue only,
        \item[**] difference $\leq$5 days,
        \item[***] difference $>$5 days.
      \end{description}
  \item[\textbf{(12) Magnitude at maximum}] If known, the magnitude at maximum is given. In cases where the ASC and SSC list the magnitude at maximum for different photometric bands, both magnitudes are given separated by ``$|$'' unless they are equal. The number of asterisks means:
      \begin{description}
        \item[*] value comes from one catalogue only,
        \item[**] difference $\leq0.5^{\rm{m}}$,
        \item[***] difference $>0.5^{\rm{m}}$.
      \end{description}
  \item[\textbf{(13) Maximum magnitude band}] Here the photometric band of the magnitude at maximum is given. An asterisk after the magnitude band means that this value was taken from one catalogue only. If the ASC and SSC list different bands, both bands are given separated by ``$|$''.
  \item[\textbf{(14) Host galaxy position}] The host galaxy's right ascension and declination in J2000.0 coordinates. An asterisk means that the host galaxy position was modified manually (see next chapter).
  \item[\textbf{(15) Host galaxy redshift}] The redshift of the host galaxy or the SN, which is only given in the ASC and SSC. There is a difference between the redshift derived from the host galaxy or SN spectrum because of the explosion velocity. For nearby galaxies this may be relevant because the velocity of the explosion can be comparable to the Hubble escape velocity. In cases where the redshift of the host galaxy is found in HyperLeda, this value is given instead of the ASC or SSC values. The number of asterisks means:
      \begin{description}
        \item[*] ASC deviates by more than 10\% from HyperLeda value,
        \item[**] SSC deviates by more than 10\% from HyperLeda value,
        \item[***] ASC and SSC deviate by more than 10\% from HyperLeda value.
      \end{description}
      In all other cases the ASC and SSC redshift values are compared. If available, the ASC or otherwise the SSC value is given. The number of asterisks means:
      \begin{description}
        \item[****] SSC within 10\% of ASC value,
        \item[*****] SSC outside 10\% of ASC value,
        \item[******] value given only in the ASC,
        \item[*******] value given only in the SSC.
      \end{description}
  \item[\textbf{(16-17) Redshift-independent distance to host galaxy}] This distance is the error-weighted mean of all available measurements in Mpc. In cases where no error on the distance measurement is given, an error of 20\% is assumed for the calculation of the mean. Column 17 gives the statistical uncertainty of the weighted mean as error.
  \item[\textbf{(18) Circulars}] This column gives a list of all circulars (CBETs and IAUCs) where this SN designation is mentioned. A number starting with ``C'' stands for a CBET and a number starting with ``I'' for an IAUC.
\end{description}

\section{Detailed unification procedure}
This section describes in detail how the information for the USC was collected and unified. As a general rule, a particular piece of information is entered into the USC if it is consistent between all catalogues.

The CBAT list usually contains the information that was available at the time of the discovery of the SN, while the other two SN catalogues may have been updated with newer information that was published later. Therefore, the CBAT information is generally considered less accurate. Unless stated otherwise, deviations in the CBAT list from a consistent value in the ASC and SSC are disregarded.
\begin{description}
  \item[\textbf{(1) Supernova designation}] Even at the first glance it becomes apparent that the input SN catalogues hold a different number of SNe (CBAT 5584, ASC 5543, SSC 5528). The four galactic SNe listed in the CBAT list and the SNe in the SSC that do not have an official designation assigned by CBAT were ignored. The CBAT list is longest because it lists objects that later turned out to be no SNe. These objects can be active galactic nuclei (AGNs, e.g. QSOs), foreground stars, or others. Another class, known as supernova imposters, are $\eta$ Carinae-like outbursts or giant eruptions of luminous blue variables (LBVs). It can be debated whether or not these objects should be listed in a SN catalogue because their luminosity can reach that of the least-luminous core-collapse SNe. The approach of the USC is to exclude all objects that are positively confirmed as no SNe.

      Table \ref{tab:Objects_not_a_SN} lists the 59 objects that were not considered for the USC. Most of them were rejected based on comments given in the CBAT list (reproduced in Table \ref{tab:Objects_not_a_SN} if no other reference is given). Most excluded objects have a comment in the CBAT list, but some are listed without any additional comment. SN1916A was excluded because there are no data in the CBAT list and this SN is not listed in the ASC or SSC. The ASC is the only catalogue that lists \object{SN1984Z}, which was therefore excluded from the USC.

      The CBAT list contains 58 of the 59 objects from Table \ref{tab:Objects_not_a_SN}. The ASC includes the 16 supernova imposters \citep[taken from][]{SN_Imposters} and \object{SN1984Z}. The SSC lists the nova \object{SN2010U} \citep{SN2010_Nova} and the dwarf nova \object{SN2005md} (see ATel \#2750) and 15 out of the 16 SN imposters. Fifteen SNe are not listed in the SSC. Except for \object{SN2010ma} and \object{SN2003ma} they are all marked uncertain or unconfirmed in the ASC.

\begin{table}[!htb]
\caption{Objects that turned out to be no supernovae and were excluded from the USC are listed with the reason for exclusion.}
\label{tab:Objects_not_a_SN}
\centering
\begin{tabular}{l p{0.7\columnwidth}}
\hline\hline
Name & Reason\\
\hline
1916A & Unconfirmed, listed in CBAT list without data\\
1950E & Astroid 2093 Genichesk\\
1954J & Supernova imposter\tablefootmark{a}\\
1956C & Astroid 9574 Taku\\
1961V & Supernova imposter\tablefootmark{a}\\
1961W & Designation ommited\\
1961X & Identical with SN 1961T\\
1967F & Uncertain nature\tablefootmark{b}\\
1973G & Foreground variable\tablefootmark{b}\\
1974L & Foreground variable\tablefootmark{b}\\
1983H & Foreground star\\
1984Z & Unconfirmed, listed only in the ASC\\
1985J & Possible CV NSV18704\tablefootmark{c}\\
1986D & H II region?\\
1986F & H II region\\
1986H & Unconfirmed\tablefootmark{d}\\
1987E & Foreground star\\
1987G & Identical with SN 1987D\\
1987H & Foreground star\\
1988X & H II region?\\
1990C & H II region\\
1991W & Foreground star\tablefootmark{e}\\
1991ap & QSO\\
1992W & Foreground M dwarf\\
1992X & Foreground M dwarf\\
1993U & QSO\\
1993V & QSO\\
1997bs & Supernova imposter\tablefootmark{a}\\
1998di & Dwarf nova\\
1998ev & Not variable\\
1999bs & Foreground variable, UGC 11093\\
1999bw & Supernova imposter\tablefootmark{a}\\
1999cu & AGN\\
1999cv & AGN\\
1999db & QSO\\
1999dc & AGN\\
2000ch & Supernova imposter\tablefootmark{a}\\
2000dh & Foreground star\\
2001ac & Supernova imposter\tablefootmark{a}\\
2001bh & Foreground star\\
2001bn & Galactic blue variable\\
2002bu & Supernova imposter\tablefootmark{a}\\
2002kg & Supernova imposter\tablefootmark{a}\\
2003aw & Dwarf nova\\
2003ec & Foreground star\\
2003gm & Supernova imposter\tablefootmark{a}\\
2003lr & Minor planets\\
2005md & Dwarf nova\tablefootmark{f}\\
2006U & AGN\\
2006bv & Supernova imposter\tablefootmark{a}\\
2006fp & Supernova imposter\tablefootmark{a}\\
2007sv & Supernova imposter\tablefootmark{a}\\
2008S & Supernova imposter\tablefootmark{a}\\
2008ft & H II region?\\
2009ip & Supernova imposter\tablefootmark{a}\\
2010U & Nova\tablefootmark{g}\\
2010da & Supernova imposter\tablefootmark{a}\\
2010db & Very red M-type star\\
2010dn & Supernova imposter\tablefootmark{a}\\
\hline
\end{tabular}
\tablefoot{
\tablefoottext{a}{\citet{SN_Imposters}}
\tablefoottext{b}{\citet{barbon1984}}
\tablefoottext{c}{IAUC 4070}
\tablefoottext{d}{IAUC 4219}
\tablefoottext{e}{IAUC 5270}
\tablefoottext{f}{Atel \#2750}
\tablefoottext{g}{\citet{SN2010_Nova}}
}
\vspace{5mm}
\end{table}

  \item[\textbf{(2) Supernova host galaxy}]
  The main challenge for unifying the host galaxy identifier is that different SN catalogues may list the same host galaxy with different identifiers. Therefore, lists of possible galaxy aliases were created from HyperLeda. The host galaxy names that are given in the different catalogues were first checked for agreement with this list. If so, the HyperLeda principal name was chosen for the USC. If the CBAT list had no host galaxy and the ASC and SSC agreed on the host galaxy that host galaxy was taken.

  Some galaxies are not listed in HyperLeda. In these cases the host galaxy was taken if at least the ASC and SSC agreed. All other cases were handled manually. They can be found in Table \ref{tab:manual_galaxy_unification}. The CBAT list always contains only the main NGC galaxy name instead of the sub-member (e.g. NGC 1573 instead of NGC 1573A for \object{SN2010X}). This and some obvious typographic errors were fixed manually and are not listed in Table \ref{tab:manual_galaxy_unification}. If a galaxy group is listed the best-fitting galaxy within that group was selected.

\begin{table}[!htp]
\caption{Manually unified host galaxies}
\label{tab:manual_galaxy_unification}
\centering
\begin{tabular}{l p{0.7\columnwidth}}
\hline\hline
Name & Reason\\
\hline
2010kr & Misidentification CBET\\
2010jl & UGC5189A not recognised by HyperLeda\\
2010af & MCG+15-01-010 is $1\arcmin$ closer to the SN than NGC3172\\
2010ad & Misidentification CBET\\
2010X & NGC1573A not recognised by HyperLeda\\
2009ij & Galaxy group\\
2009el & Galaxy group\\
2009bv & Misidentification CBET\\
2008hm & Misidentification CBET\\
2008fu & Galaxy group\\
2008bv & Galaxy group\\
2007so & Uncertain alias in the SSC\\
2007R & Galaxy group\\
2006rs & Uncertain alias in the SSC\\
2006el & Galaxy group\\
2006ej & NGC191A not recognised by HyperLeda\\
2006dn & Galaxy group\\
2006ay & UGC10116 excluded (inconsistent redshift\tablefootmark{a})\\
2005nb & Galaxy group\\
2005em & PGC1148248 is $0.5\arcmin$ closer to the SN than IC307\\
2004gu & FGC175A not recognised by HyperLeda\\
2004bt & Galaxy group\\
2003fa & Redshift error NED\tablefootmark{b}\\
2003ei & Galaxy group\\
2003H & IC 2163 is $0.3\arcmin$ closer to the SN than NGC2207 \\& (interacting galaxy pair)\\
2002jg & Galaxy group\\
2002eh & NGC917 not recognised by HyperLeda\\
2002ec & Galaxy group\\
2002dc & HDFN2-264.1 not recognised by HyperLeda\\
2002bt & Galaxy group\\
2002bn & Galaxy group\\
2001ej & Galaxy group\\
2001ck & Galaxy group\\
2000dl & Galaxy group\\
2000cp & PGC57064 is a "two bulged" galaxy\\
2000ci & Misidentification IAUC\\
1999gw & Galaxy group\\
1999dj & Galaxy group\\
1999D & Galaxy group\\
1998er & GH9-2 not recognised by HyperLeda\\
1998eh & Galaxy group\\
1997T & Galaxy group\\
1995T & Galaxy group\\
1992bb & IRAS21156-0747 not recognised by HyperLeda\\
1991bc & Galaxy group\\
1990ak & Galaxy identified manually\\
1985C & Misidentification IAUC\\
1980I & Spectral analysis favours NGC4374\tablefootmark{c}\\
1976A & ASC and SSC agree (no discovery notice)\\
1972T & Only listed in the ASC\\
1969A & Misidentification IAUC?\\
1967I & Closest galaxy to SN position\\
1965A & Closest galaxy to SN position\\
1964C & Closest galaxy to SN position\\
1961K & Closest galaxy to SN position\\
1960F & Closest galaxy to SN position\\
1958E & Only listed in the ASC\\
1956B & Closest galaxy\\
1953A & Closest galaxy to SN position\\
1921A & Antenna extending south from NGC4308\tablefootmark{d}\\
\hline
\end{tabular}
\tablefoot{
\tablefoottext{a}{CBET 489}
\tablefoottext{b}{IAUC 8146}
\tablefoottext{c}{\citet{SN_rate_elliptical}}
\tablefoottext{d}{\citet{NGC4038-4039}}
}
\end{table}
  \item[\textbf{(3) Supernova position}] For unifying the SN positions, the values from the CBAT list, the ASC and SSC were compared. Positions were considered consistent if the deviations are within 5\arcsec. If all three catalogues agreed within this accuracy, the ASC position was entered into the USC. If the CBAT list had no coordinates, it was ignored. All cases where the available positions were not consistent were checked manually. Mostly, the conflicts could be resolved, and all other cases are marked with two asterisks. For some cases the SN and host galaxy positions and offsets were fixed manually if the inconsistencies could be fixed by a specific combination of the three values, e.g.\ assuming an obvious typographic error in one catalogue. Some positions of older SNe were obtained directly from specific publications \citep{precise_astrometry,VanDyk92,VanDyk96a,VanDyk96b}. These were not compared to the positions in the other catalogues. A few obvious typographical errors were fixed by hand.
  \item[\textbf{(4) Supernova offset}] Offsets were considered to be consistent if they agreed within 5\arcsec. The USC always lists the ASC offsets, and if these were unavailable the SSC values. Those cases where the other catalogues are inconsistent were flagged. Some offsets were taken from \citet{VanDyk92} and \citet{VanDyk96a}. The SN position was then calculated with the host galaxy position and this offset.
  \item[\textbf{(5-7) Supernova type}] In cases where the SSC had no value and the ASC agreed with the CBAT list, the value from the CBAT list and the ASC was taken. Cases are flagged where only one catalogue has a value or one catalogue does not list the subtype. For all other cases the types separated by ``$|$'' are given.

      In cases where the catalogues did not agree if the SN spectrum was peculiar, the value from the ASC was taken and a flag added. The same procedure was performed for different subtypes of SN type II.
  \item[\textbf{(8-10) Discovery date, magnitude and band}]
      The unification of these three quantities depends on which photometric bands are given in the different catalogues. A flag was added if only one catalogue provides a band and both bands are given if they differ. The date and magnitude are the ASC values (if these were unavailable, the SSC values are given) unless only one catalogue has a band (then the values from this catalogue are given) or the bands are different (then both are given separated by ``$|$'' unless they are equal). The ASC replaces a possible discovery date with the date of the optical maximum, while the SSC lists both. The ASC and SSC both replace the discovery magnitude if the magnitude of the optical maximum is known. Small differences below $0.1^{\rm{m}}$ were ignored. Some obvious typographical errors were fixed by hand.
  \item[\textbf{(11-13) Maximum date, magnitude and band}]
      The unification of these values was easier because they are listed only in the ASC and SSC. Again, the information that was included in the USC depended on the available magnitude bands. A flag was added if only one catalogue provided a band and both bands are given if they are different. The date and magnitude are the ASC values (if these were unavailable the SSC values) unless only one catalogue had a band (then the values from this catalogue are given) or the bands were different (then both are given separated by ``$|$'' unless they are equal). Small differences below $0.05^{\rm{m}}$ were ignored. Some errors were again fixed by hand.
  \item[\textbf{(14) Host galaxy position}]
     The host galaxy position was taken from HyperLeda. In cases where this was unavailable, the ASC position was used.
  \item[\textbf{(16-17) Redshift-independent distance to host galaxy}]
     The redshift-independent distances were determined from the NASA/IPAC master list of galaxy distances (NED) and in a few cases from distance moduli listed in HyperLeda. The distance moduli in NED that were taken from HyperLeda were excluded to avoid dual use of these values.
\end{description}

\section{Statistical studies with the USC}
\label{sec:studies}
In this section a few observables of the USC are investigated as examples of possible statistical studies. The meta-catalogue capabilities of the USC are demonstrated by selecting a subsample of high-quality data. Uncertain SNe (followed by at least one asterisk) were excluded, which removed 428 SNe. About half of them are either old (before 1987) or have a very high redshift (higher than 0.1). The other quality cuts are described in the description of the corresponding plots.

The number of discovered SNe per year was already shown above in Fig.~\ref{fig:snperyear}. The discovery year was taken from the USC. In some cases this is not identical with the year from the SN designation (\object{SN1961U}, \object{SN1991B}, \object{SN2002lt} and \object{SN2010E}). The ASC magnitude is used for non-unified magnitudes at maximum (caused by inconsistencies or different bands). The USC includes only SNe with an official designation assigned by CBAT. About 300 SNe from high-redshift surveys are not covered \citep{HST,SNLS}.

\begin{figure}[!htp]
\includegraphics*[width=\columnwidth]{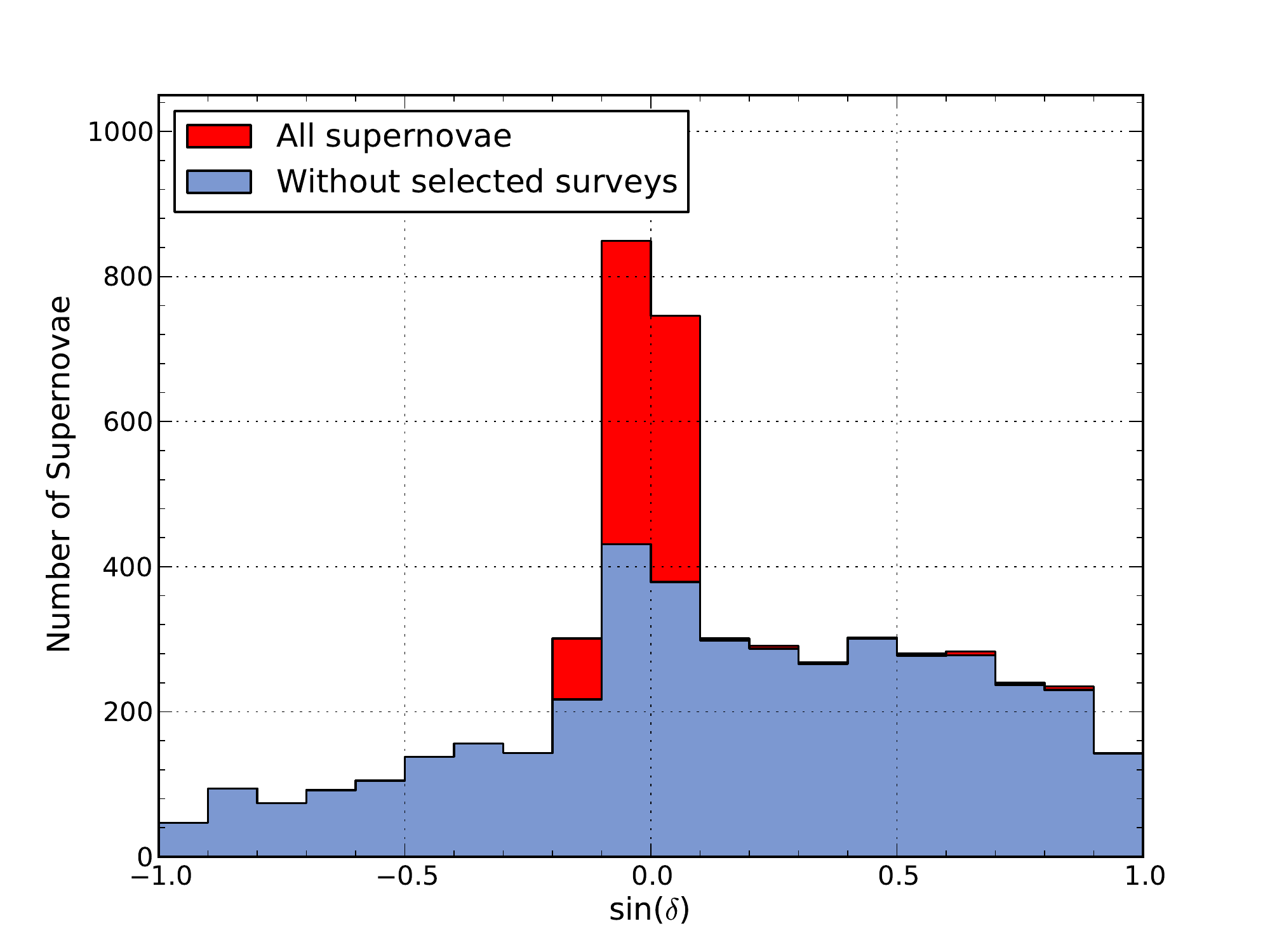}
\includegraphics*[width=\columnwidth]{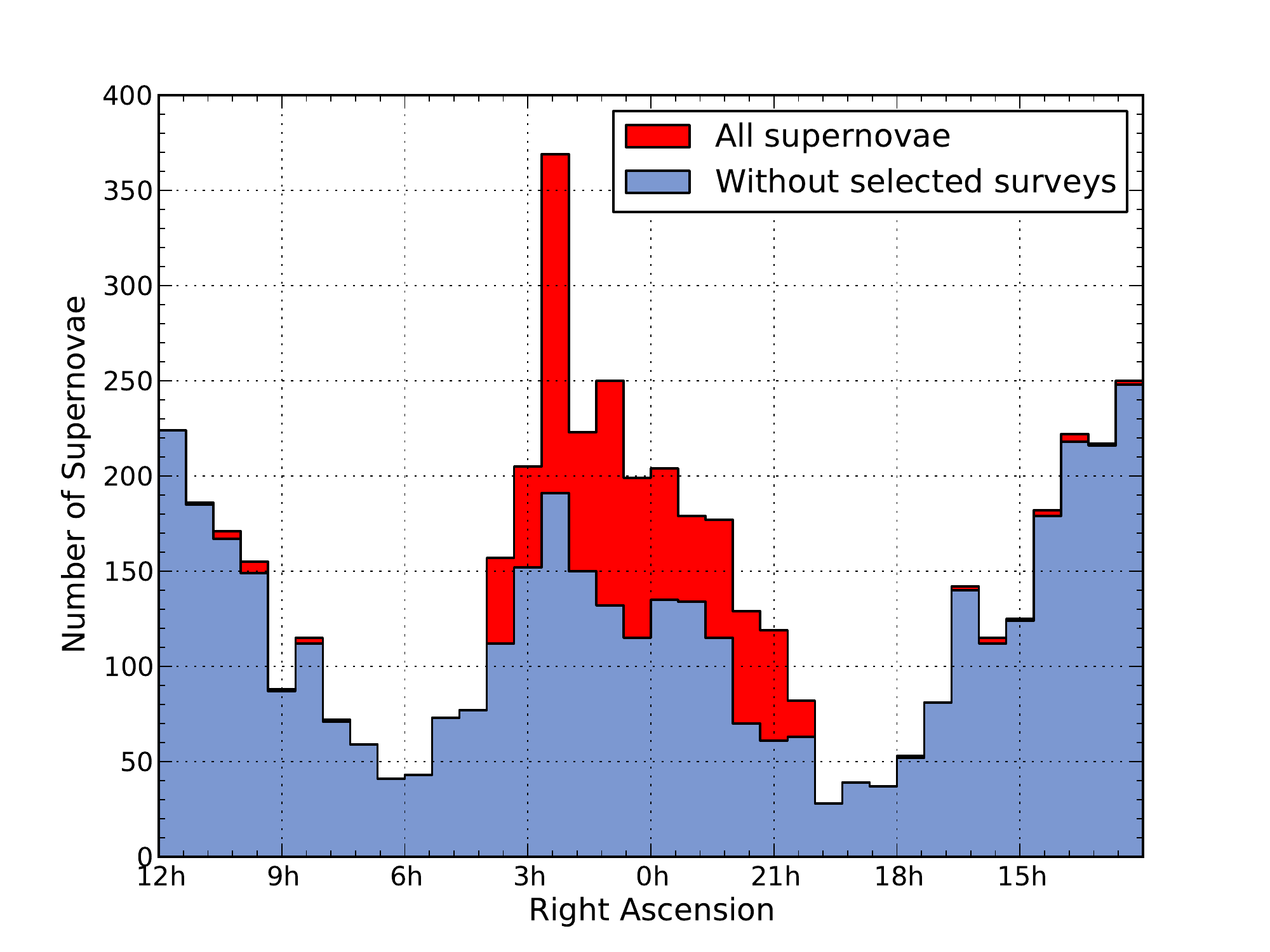}
\caption{Distributions of declination (top) and right ascension (bottom) for all SNe with a precise position. The contributions from SNe discovered during the selected SN surveys are indicated by the darker coloured area.}
\label{fig:decra}
\end{figure}

For the following positional plots all SNe with a remaining positional uncertainty larger than 5\arcsec after unification were excluded. This corresponds to the removal of all entries with one asterisk and excludes five SNe. The distributions of declination and right ascension are shown in Fig.~\ref{fig:decra}. Both distributions are not flat, but reflect different observational biases for different regions of the sky. The distribution of declination shows a clear north-south asymmetry, corresponding to different numbers of observations of the northern and southern sky. The peak around the equator is caused by various SN surveys. Equatorial observations have the advantage that the discovered SNe are accessible to telescopes in both hemispheres for spectral follow-up observations. As an example the distribution of the Sloan Digital Sky Survey \citep[SDSS;][]{SDSS} and ESSENCE \citep[Equation of State: SupErNovae trace Cosmic Expansion;][]{ESSENCE} are marked separately in the darker coloured area. Surveys mostly target at high-redshift and preferably distant SNe are found. The structure in the distribution of right ascension can be mainly attributed to obscuration by the Galactic plane, which greatly reduces the probability to detect a SN.

\begin{figure}[!htp]
\includegraphics*[width=\columnwidth]{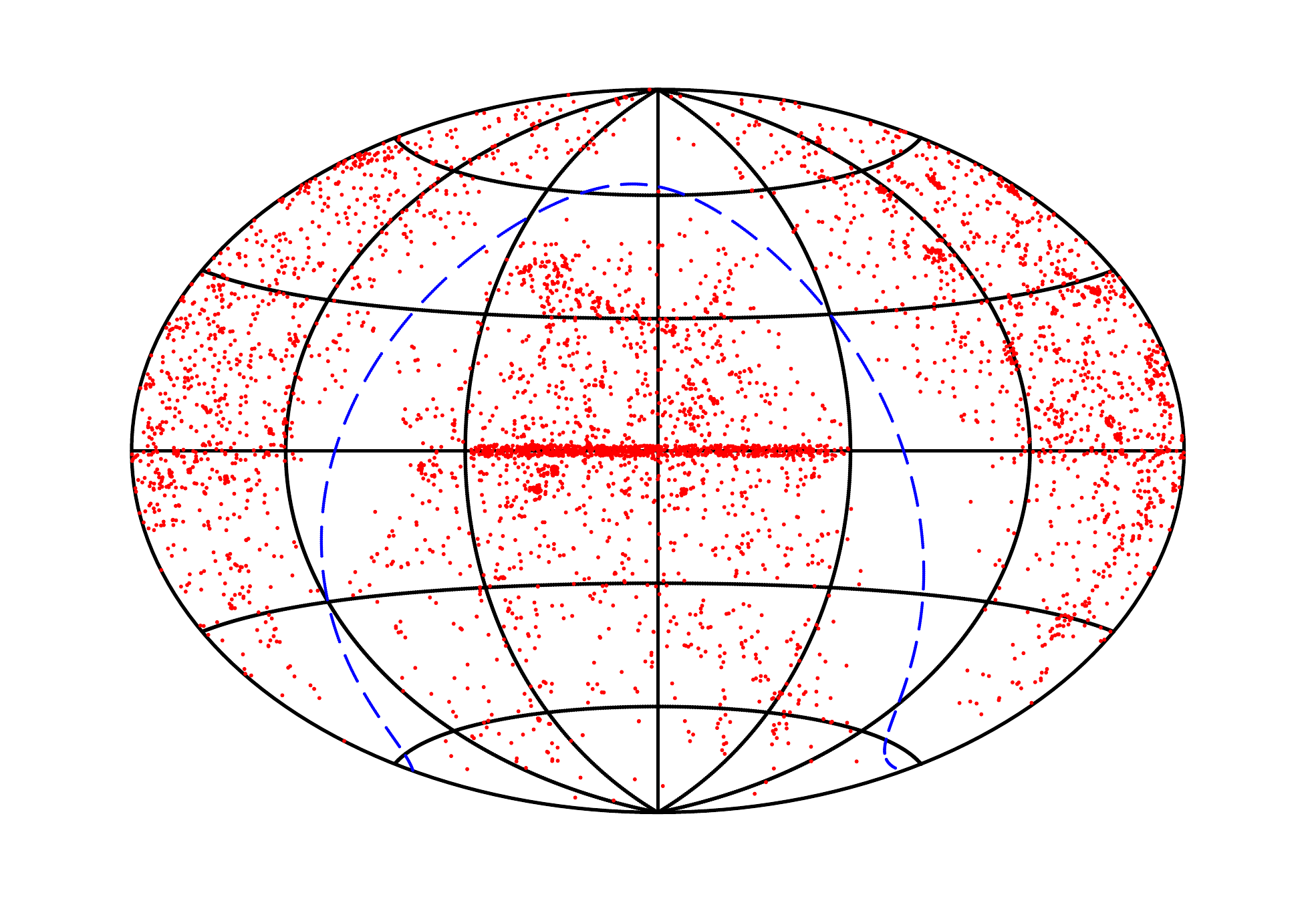}
\caption{Equatorial Hammer projection of the discovered SNe with the Galactic plane indicated by a dashed line. The x-axis for the right ascension is the same as in Fig.~\ref{fig:decra} (using the astronomical convention).}
\label{fig:radecMap}
\end{figure}

Figure \ref{fig:radecMap} shows an equatorial skymap of the discovered SNe. The contribution of surveys around the equator becomes very obvious, especially between -3h and +3h. The Galactic plane, indicated by the dashed line, clearly reduces the amount of SNe detections.

\begin{figure}[!htp]
\includegraphics*[width=\columnwidth]{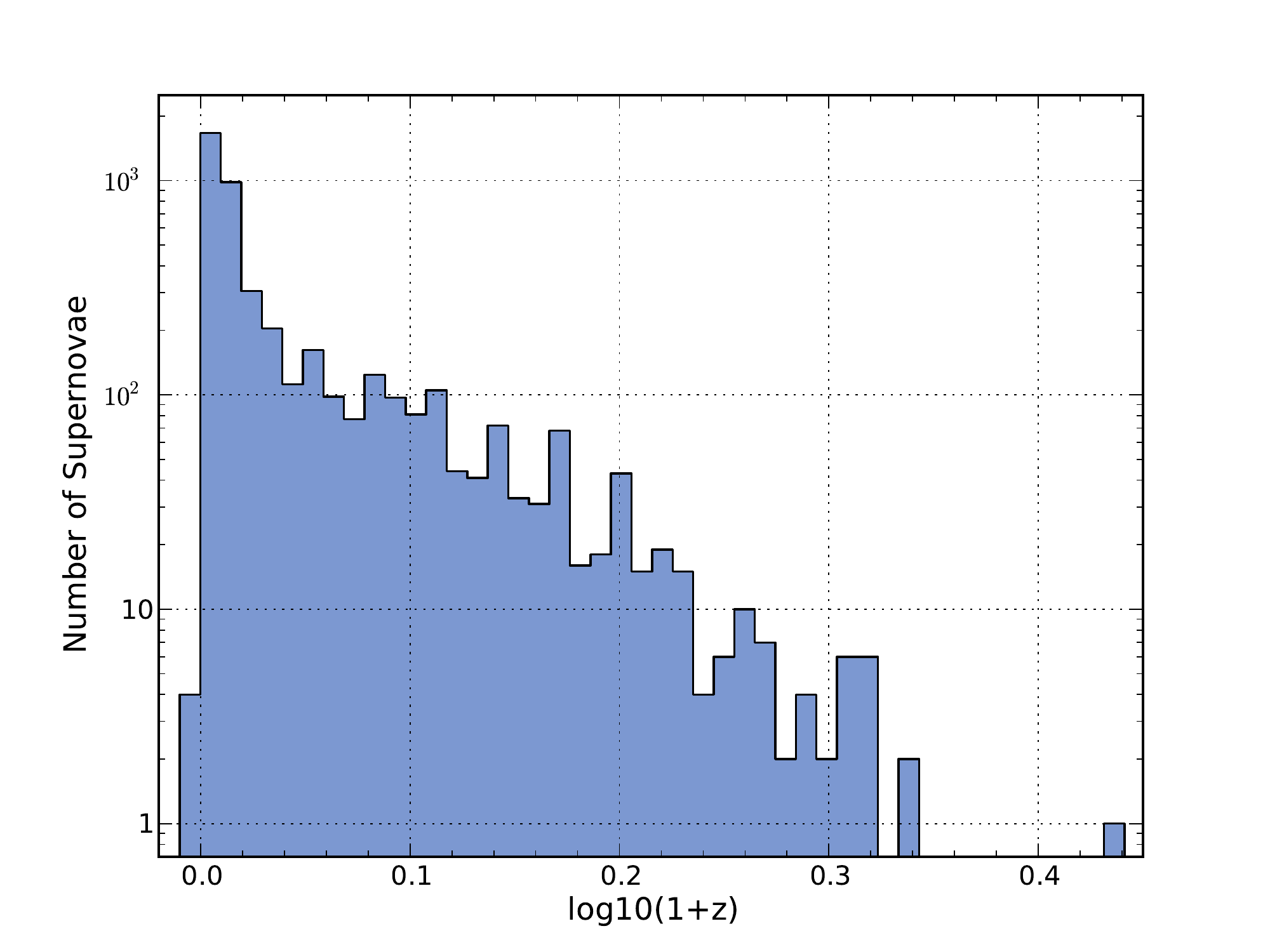}
\includegraphics*[width=\columnwidth]{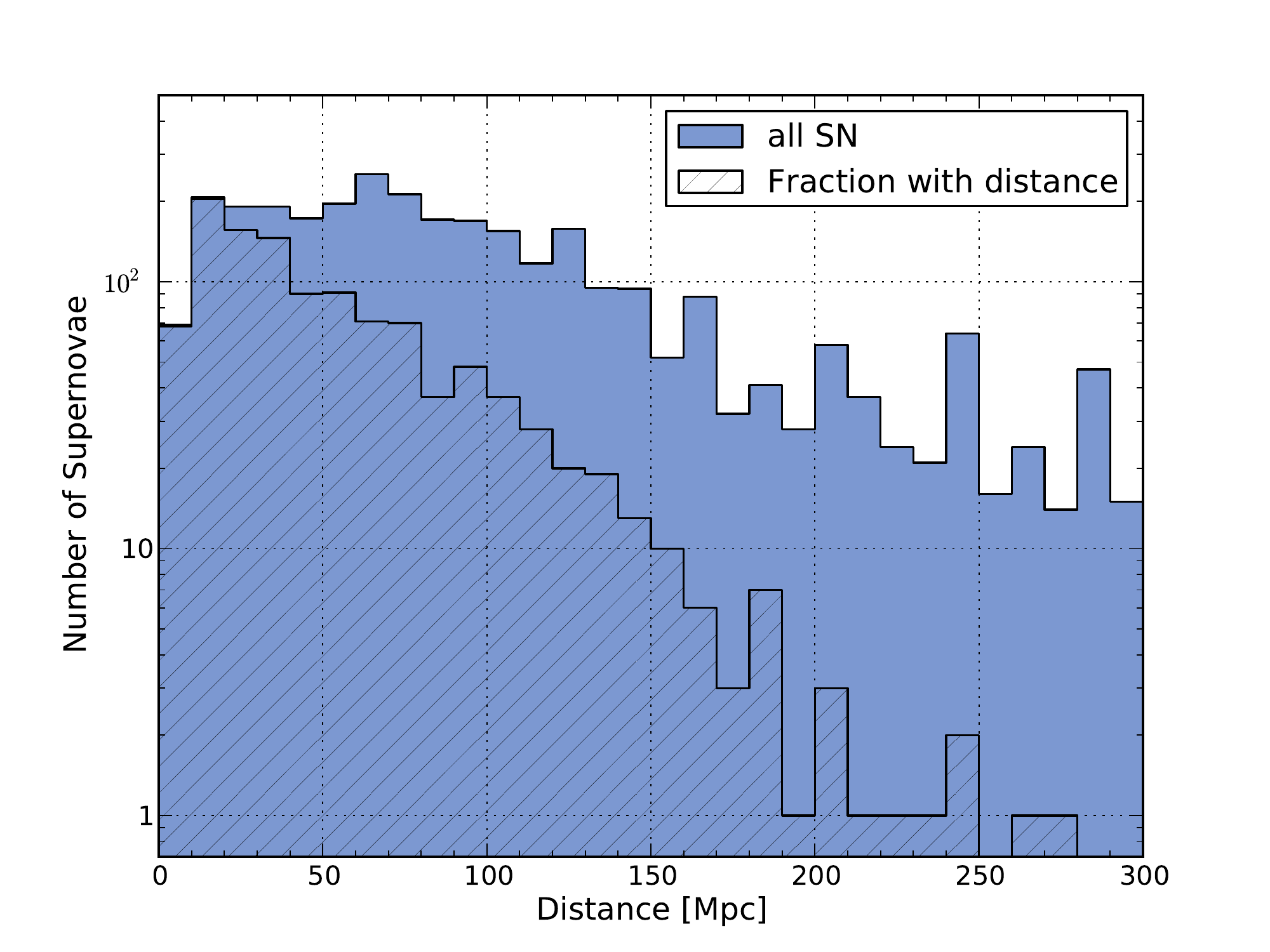}
\caption{Distributions of redshift (top) and estimated distance (bottom) for close SNe. In the bottom panel the distance is the redshift-independent one from the USC or if that was not available it was estimated using Hubble's law ($H_0=74.2 \mathrm{km\,s^{-1}\,Mpc^{-1}} $). The hatched distribution displays the fraction where a redshift-independent distance estimator was used.}
\label{fig:dist}
\end{figure}

The number of SNe as a function of redshift and distance is shown in Fig.~\ref{fig:dist}. The redshift was restricted to the HyperLeda values and to those cases for which both the ASC and SSC have a redshift that agrees within 10\% (this corresponds to the exclusion of entries with more than four asterisks). From the top plot it can be seen that the USC covers nearby up to distant SNe with redshifts above 1. Furthermore, low-redshift SNe outnumber the trend seen at high redshifts. A plausible explanation is the higher exposure time for nearby galaxies, e.g. by amateur astronomers. In the bottom plot it can be seen that the fraction of SNe with a redshift-independent distance estimate decreases with distance. This is expected because of the decreasing probability to find a reliable distance estimate for that particular host galaxy. For distances up to several tens of Mpc a redshift-independent distance estimate is obtained for a large part of SNe. The peaks in the distance distribution are related to the uncertainty that arises from the rounding of the redshift (e.g. $z=0.05$ corresponds to the peak at $\simeq 200$ Mpc). For two SNe with negative redshift no redshift-independent distance was found (\object{SN2010hc} and \object{SN2002dt}).

\begin{figure}[!htp]
\includegraphics*[width=\columnwidth]{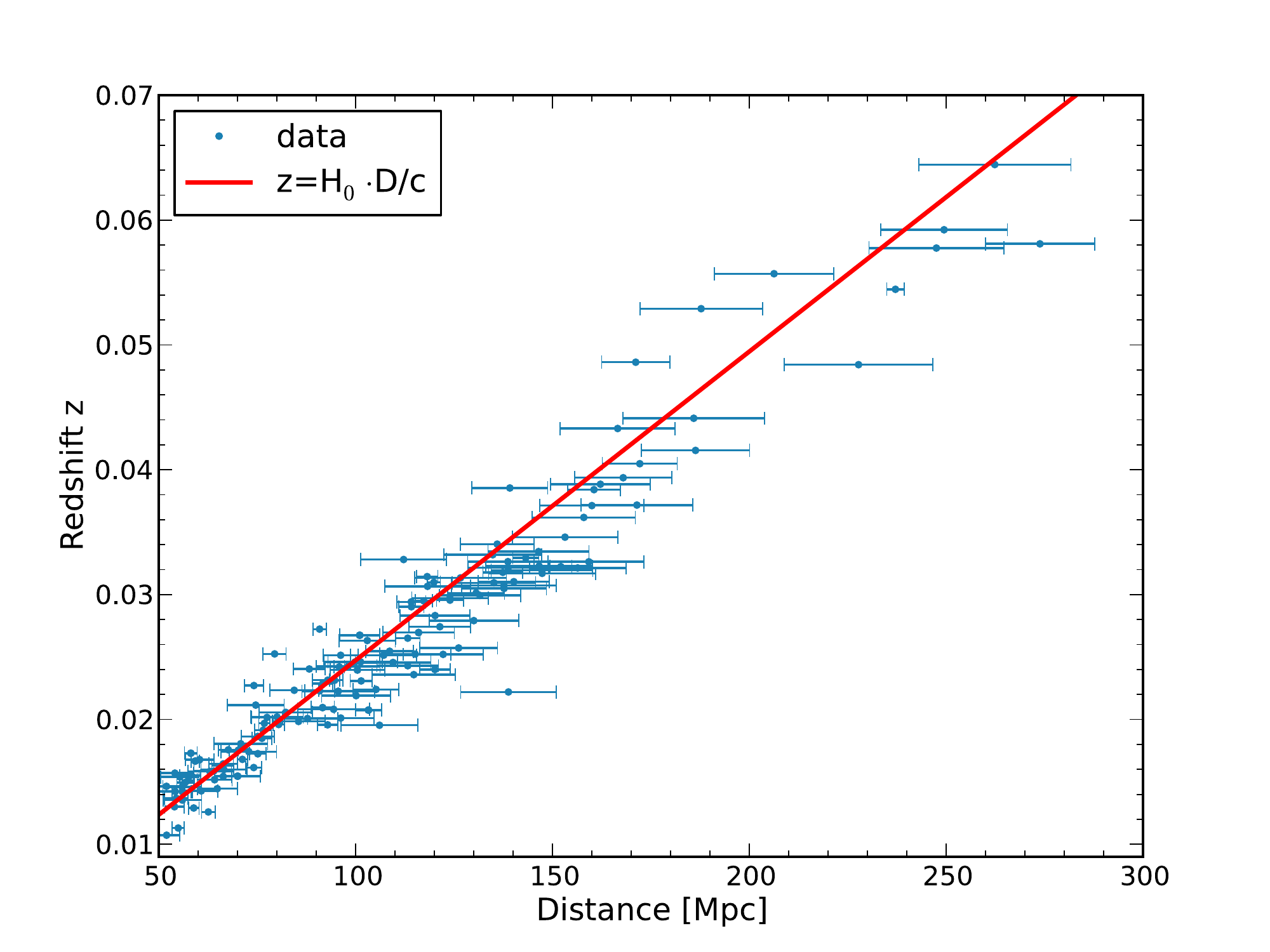}
\includegraphics*[width=\columnwidth]{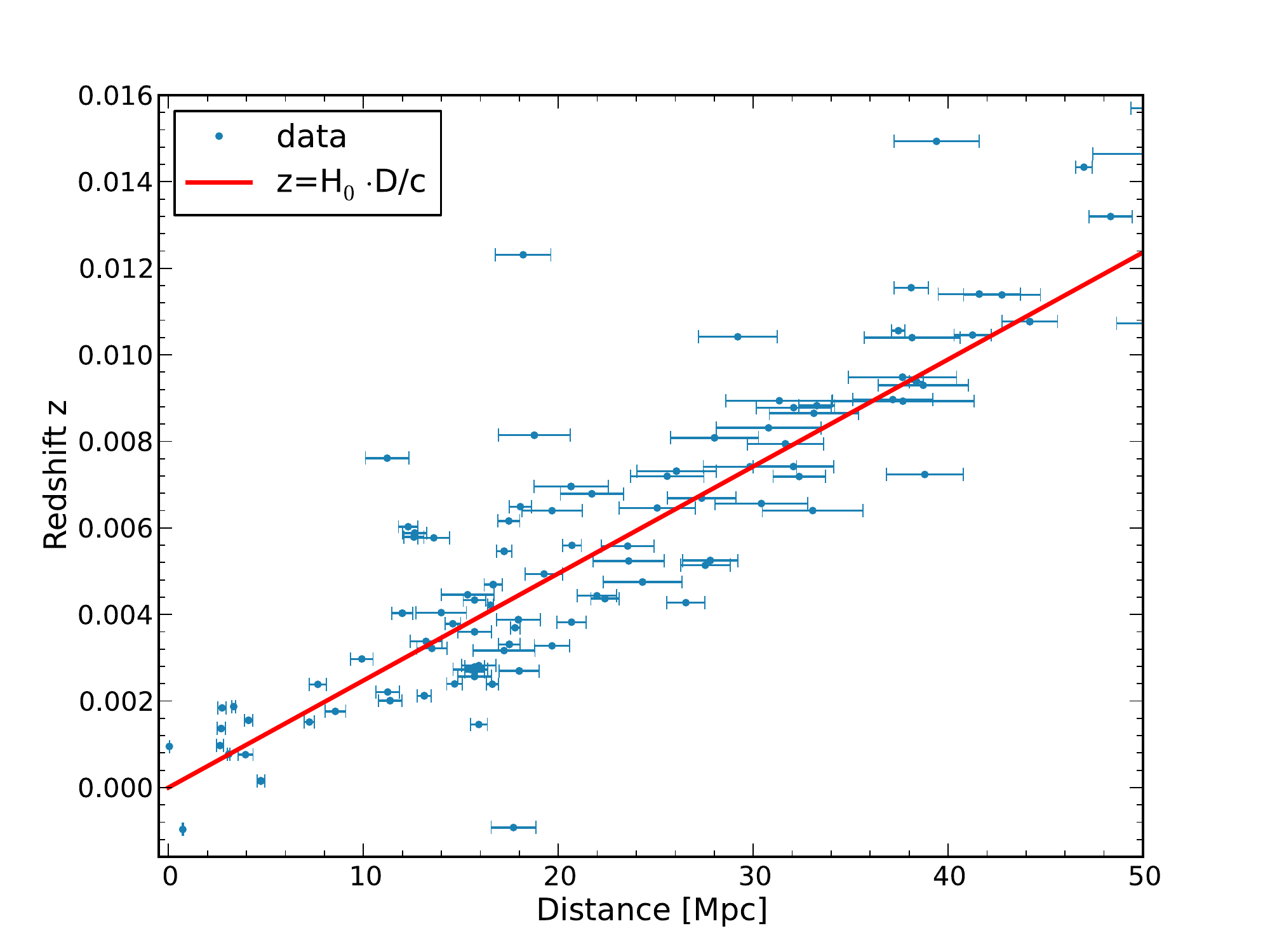}
\caption{Redshift of the SN host galaxy versus redshift-independent distance to the host galaxy. The red line corresponds to the linear Hubble law using $H_0=74.2 \mathrm{km\,s^{-1}\,Mpc^{-1}} $. Only distances with an absolute error smaller than 10\% were used.}
\label{fig:zversd}
\end{figure}

Figure \ref{fig:zversd} shows the correlation of the redshift-independent distance estimate with the host galaxy distance derived from its redshift. Only redshifts from HyperLeda are used (excluding entries with more than three asterisks) to make sure that the calculated distance corresponds to the host galaxy distance. To make the plot less crowded, only distances for which the error on the distance is smaller than 10\% were used. Evidently, particular for close SNe the redshift-independent distance estimate can deviate substantially from a redshift-based estimate, as expected from the local flow of galaxies.

\begin{figure}[!htp]
\includegraphics*[width=\columnwidth]{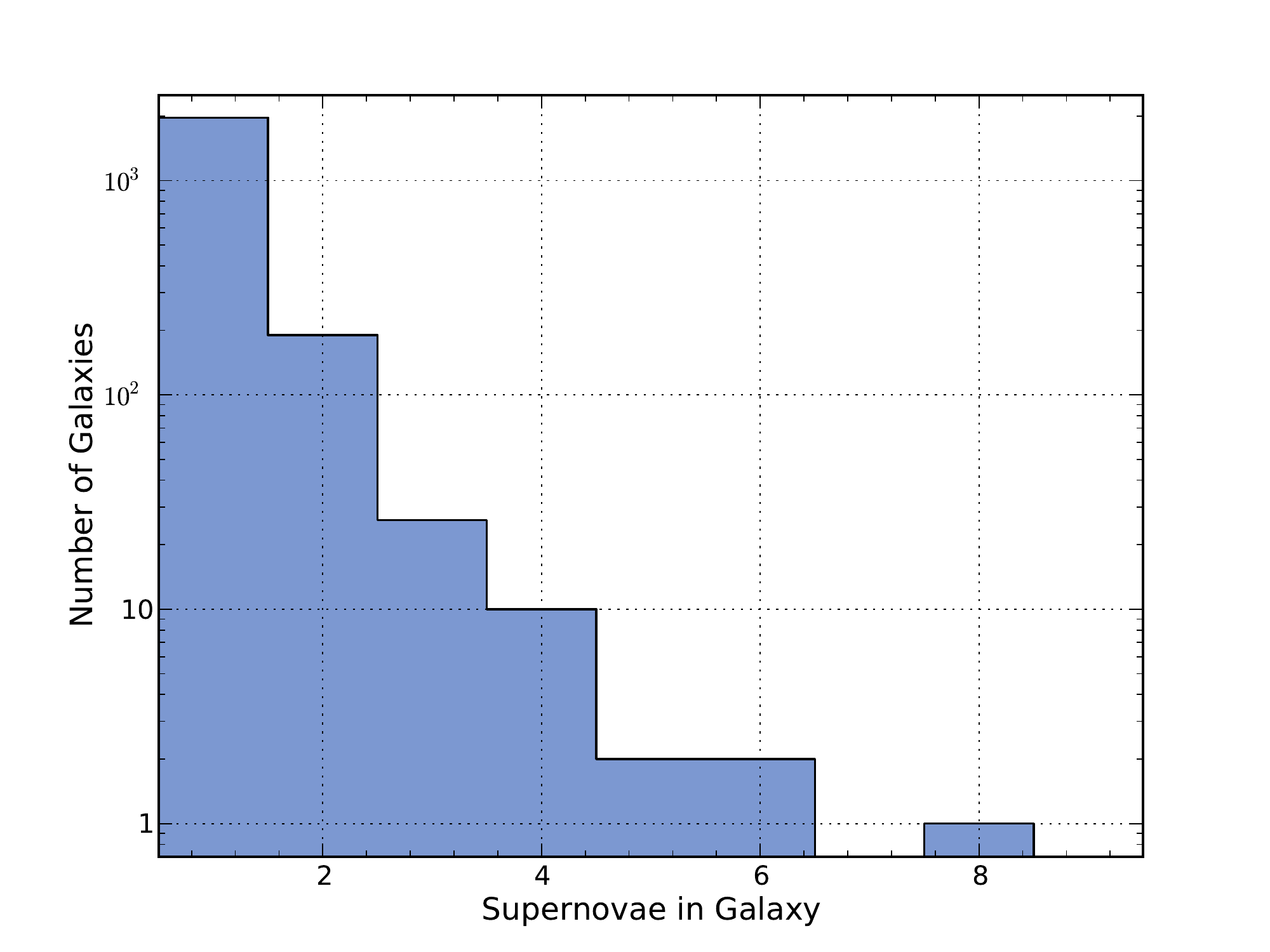}
\caption{Distribution of the number of SNe in the USC per host galaxy.}
\label{fig:snpergalax}
\end{figure}

Figure \ref{fig:snpergalax} shows the distribution of the number of SNe per identified host galaxy. More than 100 host galaxies contribute more than one SN to the USC. Most remarkable is NGC 6946 with eight observed SNe, which is therefore also named \emph{firework galaxy}.

\begin{figure}[!htp]
\includegraphics*[width=\columnwidth]{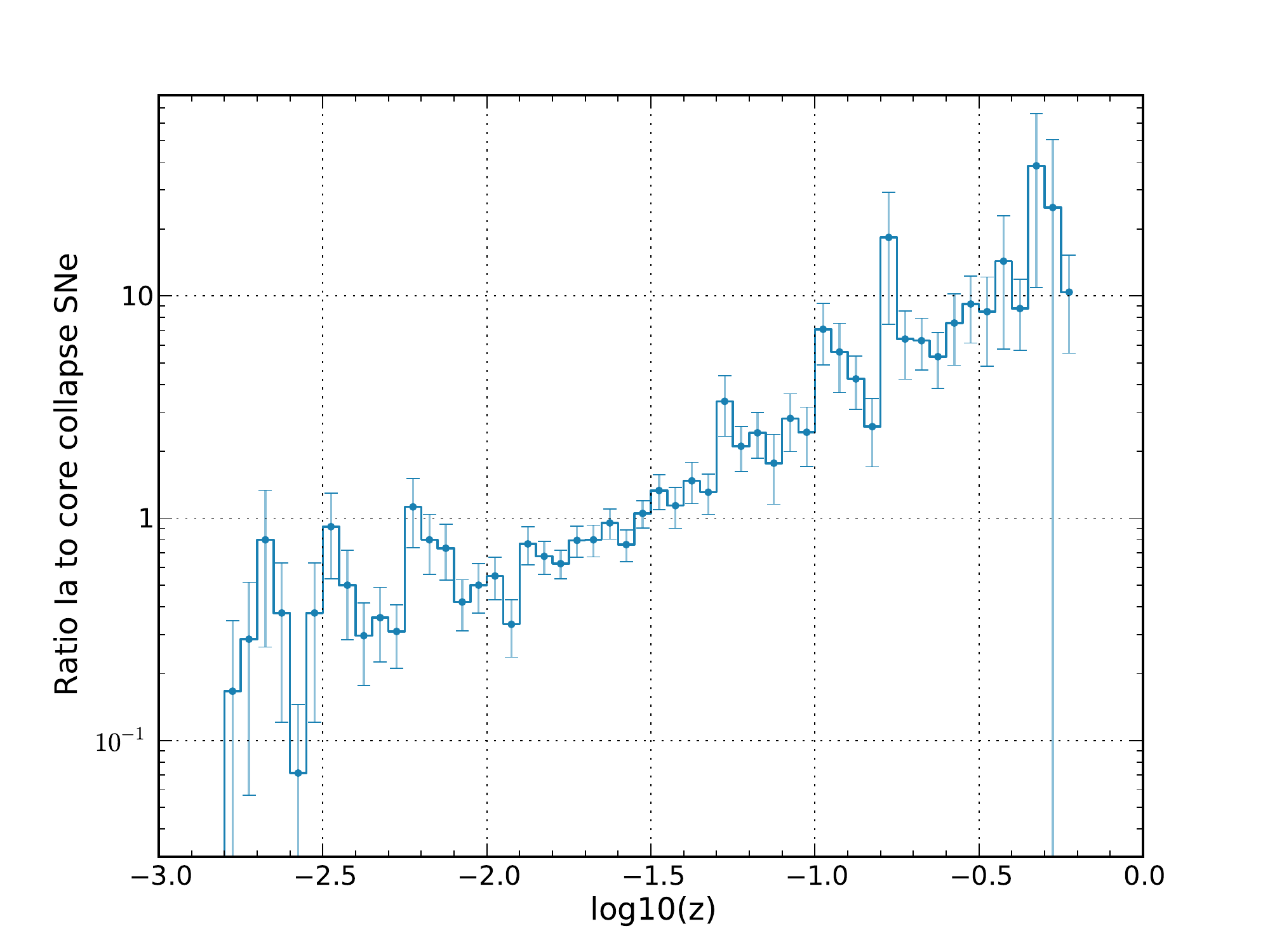}
\caption{Ratio between type Ia and core-collapse SNe as a function of redshift.}
\label{fig:sn1adist}
\end{figure}

Figure \ref{fig:sn1adist} shows the ratio between SNe of type Ia and core-collapse SNe as a function of the host galaxy or SN redshift. Non-unified types were added only if all catalogues list a type and agree that it was not type Ia or core-collapse (accepting e.g. Ib\textbar II\textbar II). The redshift was again restricted to entries with less than four asterisks. For close SNe this ratio is approximately constant around $\simeq 0.4$, but then rises linearly to about $\simeq 10$. SNe of type Ia are brighter and can therefore be identified out to farther distances. Furthermore, the SN surveys mostly target this type of SN because they are very interesting for cosmology. The spectrum that would identify a core-collapse SN is mostly not taken because of the huge amount of observation time needed for high-redshift objects. The break in the ratio at $\log\left(z\right)=-2.0$ indicates the completeness of the catalogue (about $\simeq 50$\,Mpc).

\begin{figure}[!htp]
\includegraphics*[width=\columnwidth]{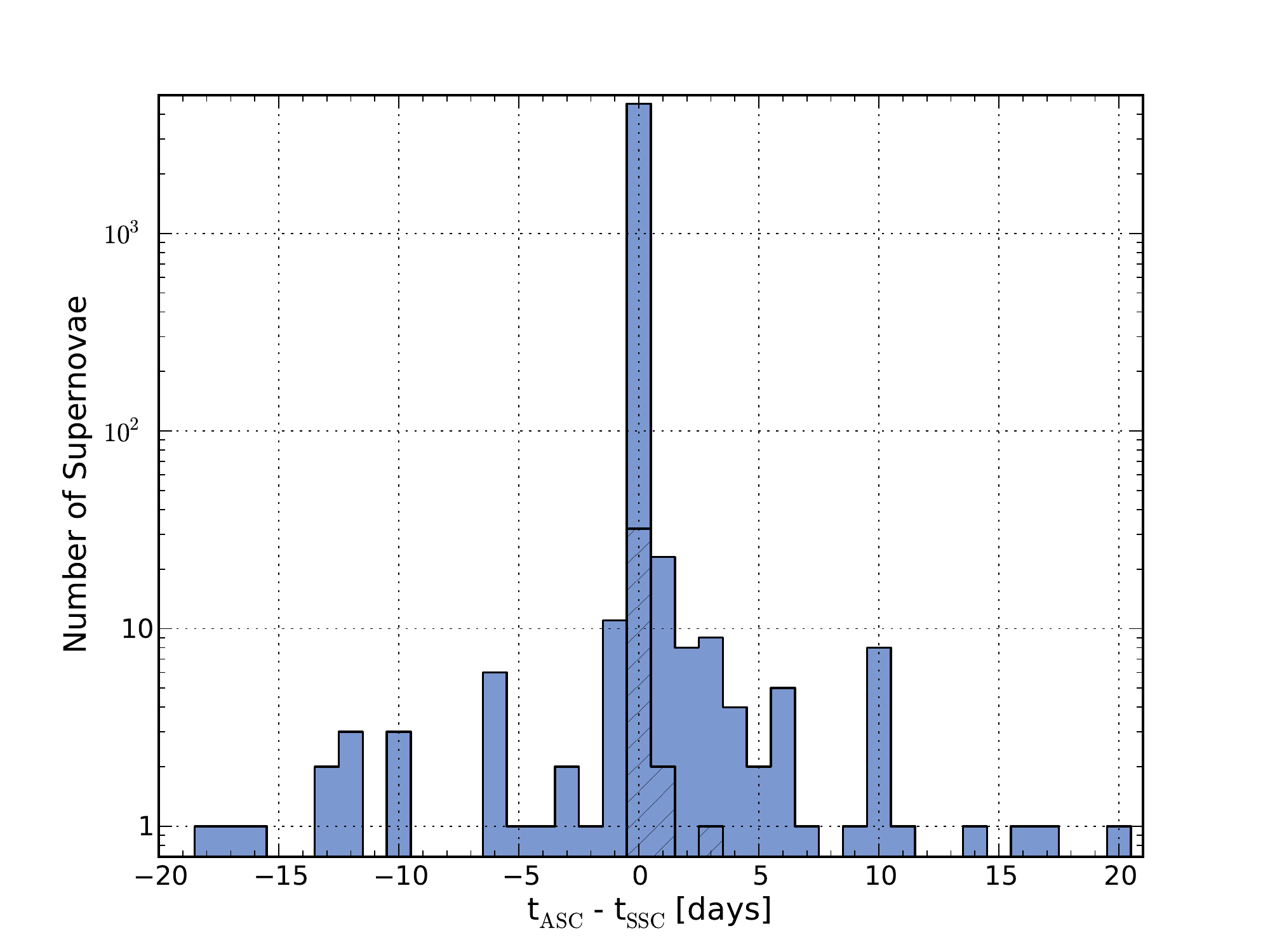}
\caption{Distribution of the difference in discovery date between the ASC and SSC. The fraction where the observational bands are not the same in the catalogues are shown as hatched distribution.}
\label{fig:diffdiscovery}
\end{figure}

The need for a meta-catalogue is demonstrated in Fig.~\ref{fig:diffdiscovery}. It shows the difference in discovery date between the ASC and SSC. In most cases they agree, but occasionally the disagreement is as big as several days. The hatched distribution shows that the majority of these differences cannot be attributed to different observation bands. A plausible explanation could be the ambiguity between the first official discovery announcement of a SN and the earliest detection, which might be discovered after official announcement.

\begin{figure}[!htp]
\includegraphics*[width=\columnwidth]{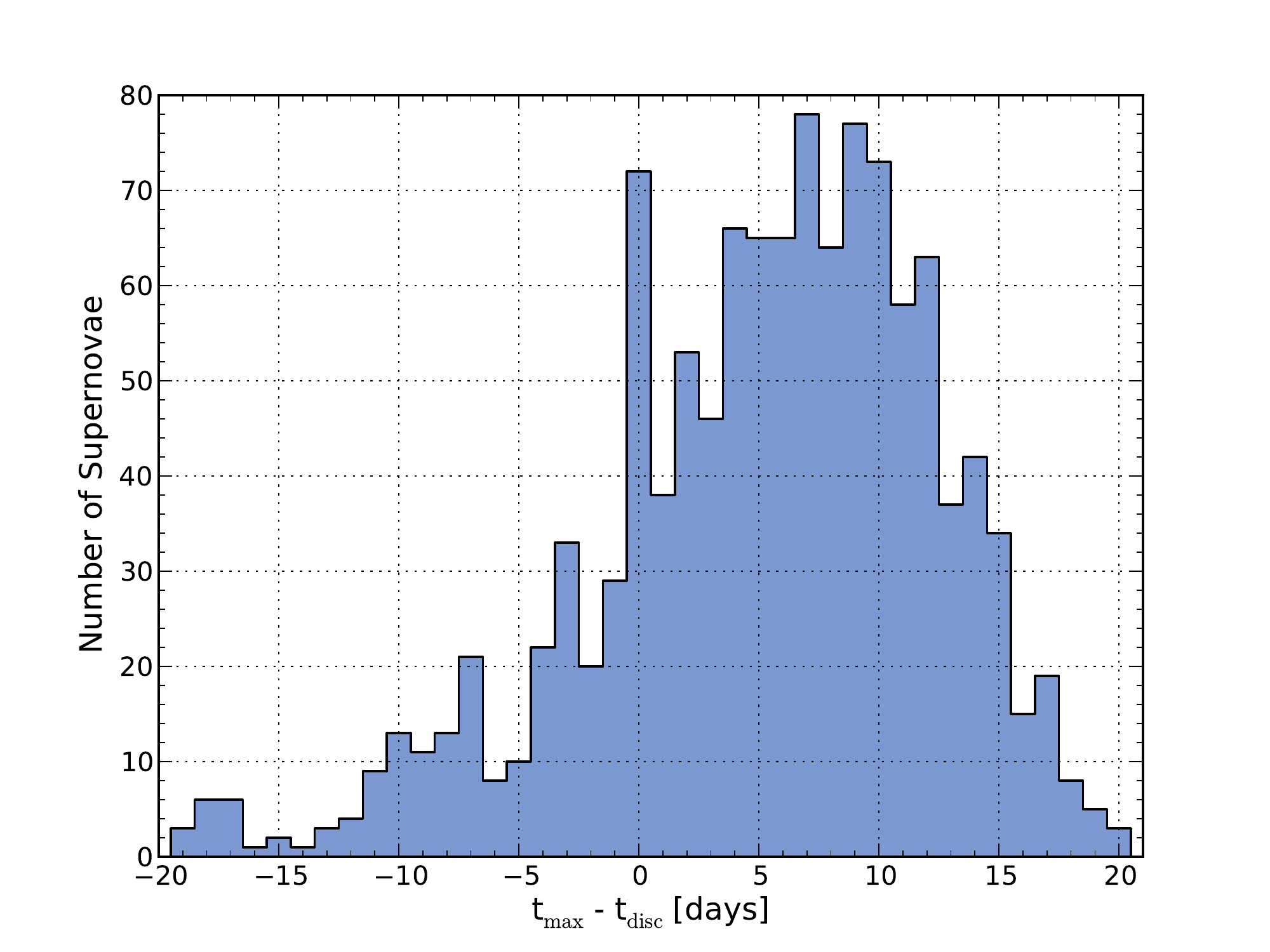}
\caption{Distribution of the differences of the date of the maximum and the discovery date for SNe with unambiguous dates. Differences of more than 20 days are excluded from this plot.}
\label{fig:diffdiscmax}
\end{figure}

The time difference between the maximum and the discovery date is shown in Fig.~\ref{fig:diffdiscmax}. The optical maximum is much more likely to be observed if the SN was discovered before. In some cases the date of the maximum can be extrapolated back or recovered based on older observations. From Fig.~\ref{fig:diffdiscmax} it can be seen that it is very unlikely that a SN is detected more than 10 days after optical maximum. The largest observed difference is \object{SN1997ab} with $t_{\rm{max}} -t_{\rm{disc}} = -323$ days \citep{SN1997ab}.

\section{Summary}
In this paper a new unified catalogue of three existing supernova catalogues is presented. During the unification procedure several inconsistencies between the catalogues were identified and errors corrected. Remaining inconsistencies are transparently marked and enable the user to select high-quality subsamples. Wherever possible, redshift-independent distance estimates were added to provide a more realistic distance than the redshift only. Several examples for statistical studies that make use of the USC capabilities are shown.

The corrected and extended information contained in this unified catalogue is intended to improve the use of SN-related information, e.g. in SN-related analyses in astro-particle physics. An example is \citet{arxiv} which, initiated this work.

\begin{acknowledgements}
We acknowledge the usage of the HyperLeda database (http://leda.univ-lyon1.fr). This research has made use of the NASA/IPAC Extragalactic Database (NED) which is operated by the Jet Propulsion Laboratory, California Institute of Technology, under contract with the National Aeronautics and Space Administration. The authors would like to thank Roberto Barbon for numerous very fruitful correspondence. Furthermore, we thank Barry F. Madore and Ian Steer from NED, Dmitry Makarov and Chantal Petit from HyperLeda, Dan Green from CBAT and Dmitry Tsvetkov and Oleg Bartunov from SAI.
\end{acknowledgements}

\bibliographystyle{aa}

\begin{appendix}

\section{Example page of the USC \label{tab:exam}}

\onecolumn
\begin{landscape}
\fontsize{2.7mm}{1mm}
\selectfont
\setlength{\tabcolsep}{1.5pt}
\LTcapwidth=\textwidth
\begin{longtable}{ccccccccccccccccccc}
\caption{An excerpt of the USC. Column 18 from the catalogue was split into two columns (CBET, IAUC) and column 16-17 joined for clarity.}\\
\hline\hline
Designation & Host galaxy & \linebreakcell[t]{SN ra \\dec} & Offset & Type & Pec. & Subtype & \linebreakcell[t]{Discov. \\date} & \linebreakcell[t]{Discov. \\mag} & \linebreakcell[t]{Discov. \\band} & \linebreakcell[t]{Max. \\date} & \linebreakcell[t]{Max. \\mag} & \linebreakcell[t]{Max. \\band} & \linebreakcell[t]{Host galaxy ra \\dec} & Redshift & D [Mpc] & CBET & IAUC\\
\hline\\
2005nc & (GRB050525) & \linebreakcell[t]{18 32 32.58 \\+26 20 22.6 } & --- & Ic* &&& 2005-05-30** & 23.96 & R* & --- & --- & --- & \linebreakcell[t]{18 32 32 \\+26 20 22 } & 0.61**** & --- & --- & 8696 & \\
2005nb & PGC39014 & \linebreakcell[t]{12 13 37.61 \\+16 07 16.2 } & \linebreakcell[t]{1.5W\\5.N} & Ic &&& 2005-12-17 & 17.2 & --- & --- & --- & --- & \linebreakcell[t]{12 13 37.68 \\+16 07 10.2 } & 0.024 & --- & 352, 357 & 8657 & \\
2005na & UGC3634 & \linebreakcell[t]{07 01 36.62 \\+14 07 59.7 } & \linebreakcell[t]{1.6W\\7.4S} & Ia &&& 2005-12-31 & 16.1* & --- & 2006-01-04* & 15.6* & --- & \linebreakcell[t]{07 01 36.84 \\+14 08 06.9 } & 0.026 & 103$\pm$7 & 350, 351 & 8655 & \\
2005mz & NGC1275 & \linebreakcell[t]{03 19 49.88 \\+41 30 18.6 } & \linebreakcell[t]{19.2E\\23.6S} & Ia &&& 2005-12-31 & 18.2* & --- & 2006-01-06* & 18* & --- & \linebreakcell[t]{03 19 48.26 \\+41 30 41.4 } & 0.018 & 64$\pm$6 & 347, 351 & 8655 & \\
2005my & E302-27 & \linebreakcell[t]{04 01 53.13 \\-41 56 08.3 } & \linebreakcell[t]{15.9E\\23.0N} & II &&& 2005-12-30 & 17.8** & --- & --- & --- & --- & \linebreakcell[t]{04 01 51.75 \\-41 56 30.2 } & 0.015 & 57$\pm$11 & 346, 358 & 8655, 8689 & \\
2005mx & --- & \linebreakcell[t]{21 40 08.2 \\+10 39 14.0 } & --- & II &&& 2005-07-01 & 19.4 & R & --- & --- & --- & \linebreakcell[t]{21 40 08 \\+10 39 14 } & 0.032******* & --- & 344 & 8651 & \\
2005mw & --- & \linebreakcell[t]{17 37 44.2 \\+11 09 03.7 } & --- & II &&& 2005-06-25 & 19.2 & R & --- & --- & --- & \linebreakcell[t]{17 37 44 \\+11 09 03 } & 0.016******* & --- & 344 & 8651 & \\
2005mv & --- & \linebreakcell[t]{20 45 16.8 \\-03 49 35.7 } & --- & Ia &&& 2005-06-21 & 19.6 & R & --- & --- & --- & \linebreakcell[t]{20 45 16 \\-03 49 35 } & 0.11******* & --- & 344 & 8651 & \\
2005mu & (PGC134323) & \linebreakcell[t]{21 35 52.5 \\-26 27 03.5 } & \linebreakcell[t]{0.4W\\1.6N}** & Ia &&& 2005-05-19 & 17.4 & R* & --- & --- & --- & \linebreakcell[t]{21 35 52 \\-26 27 03 } & 0.03******* & --- & 344 & 8651 & \\
2005mt & (PGC125573) & \linebreakcell[t]{10 24 12.9 \\-03 44 50.6 } & \linebreakcell[t]{6W\\0.4N}** & II && n & 2005-02-03 & 18.0 & R* & --- & --- & --- & \linebreakcell[t]{10 24 12 \\-03 44 50 } & 0.031******* & --- & 344 & 8651 & \\
2005ms & UGC4614 & \linebreakcell[t]{08 49 14.34 \\+36 07 47.9 } & \linebreakcell[t]{25.0W\\35.9N} & Ia &&& 2005-12-27 & 17.9*** & --- & --- & --- & --- & \linebreakcell[t]{08 49 16.41 \\+36 07 11.6 } & 0.025 & 122$\pm$10 & 343, 345 & 8651, 8652 & \\
2005mr & (GOODSJ123632.48+621513.6) & \linebreakcell[t]{12 36 32.25 \\+62 15 14.5 } & \linebreakcell[t]{1.54W\\}*** & Ia* &&& 2005-12-13 & 24.2*** & --- & --- & --- & --- & \linebreakcell[t]{12 36 32 \\+62 15 14 } & 0.68**** & --- & 340 & 8651 & \\
2005mq & --- & \linebreakcell[t]{23 20 21.79 \\-00 20 59.6 } & --- & Ia &&& 2005-11-27 & 22.4* & g* & 2005-12-08* & 22.2* & g* & \linebreakcell[t]{23 20 21 \\-00 20 59 } & 0.35**** & --- & 339 & 8651 & \\
2005mp & --- & \linebreakcell[t]{01 04 45.68 \\+00 03 20.3 } & --- & Ia &&& 2005-11-24 & 22.3* & g* & 2005-12-05* & 21.8* & g* & \linebreakcell[t]{01 04 45 \\+00 03 20 } & 0.27**** & --- & 339 & --- & \\
2005mo & --- & \linebreakcell[t]{03 50 12.90 \\-00 14 24.8 } & --- & Ia &&& 2005-11-23 & 22.5* & g* & 2005-11-16* & 22.2* & g* & \linebreakcell[t]{03 50 12 \\-00 14 24 } & 0.28**** & --- & 339 & --- & \\
2005mn & (APMUKS(BJ)B034645.37-5036.3) & \linebreakcell[t]{03 49 18.44 \\-00 41 31.4 } & --- & Ib &&& 2005-11-23 & 21.2* & g* & 2005-12-05* & 20.2* & g* & \linebreakcell[t]{03 49 18 \\-00 41 31 } & 0.05**** & --- & 339 & 8651 & \\
2005mm & --- & \linebreakcell[t]{00 13 09.55 \\+01 08 43.9 } & --- & Ia &&& 2005-11-20 & 22.2* & g* & 2005-11-12* & 22.3* & g* & \linebreakcell[t]{00 13 09 \\+01 08 43 } & 0.38**** & --- & 339 & --- & \\
2005ml & (APMUKS(BJ)B224007.45+3442.0) & \linebreakcell[t]{02 14 04.42 \\-00 14 21.1 } & --- & Ia &&& 2005-11-14 & 20.5* & g* & 2005-11-26* & 19* & g* & \linebreakcell[t]{02 14 04 \\-00 14 21 } & 0.12**** & --- & 339 & --- & \\
2005mk & --- & \linebreakcell[t]{22 42 40.64 \\+00 49 58.9 } & --- & II &&& 2005-11-08 & 21.9* & g* & 2005-11-16* & 21.1* & g* & \linebreakcell[t]{22 42 40 \\+00 49 58 } & 0.15**** & --- & 339 & 8651 & \\
2005mj & --- & \linebreakcell[t]{21 31 49.44 \\-01 04 09.8 } & --- & II* && n & 2005-11-03 & 22.5* & g* & 2005-11-16* & 21.4* & g* & \linebreakcell[t]{21 31 49 \\-01 04 09 } & 0.21**** & --- & 339 & 8651 & \\
2005mi & (APMUKS(BJ)B221828.41-10001.8) & \linebreakcell[t]{22 21 02.65 \\-00 44 53.4 } & --- & Ia &&& 2005-11-01 & 22.6* & g* & 2005-11-12* & 20.9* & g* & \linebreakcell[t]{22 21 02 \\-00 44 53 } & 0.22**** & --- & 339 & --- & \\
2005mh & --- & \linebreakcell[t]{02 44 56.68 \\+00 12 12.9 } & --- & Ia* &&& 2005-10-31 & 23.1* & g* & 2005-11-08* & 22.4* & g* & \linebreakcell[t]{02 44 56 \\+00 12 12 } & 0.39**** & --- & 339 & 8651 & \\
2005mg & UGC155 & \linebreakcell[t]{00 16 44.21 \\+07 04 19.8 } & \linebreakcell[t]{0.3E\\14.7S} & II &&& 2005-12-27 & 16.2*** & --- & --- & --- & --- & \linebreakcell[t]{00 16 44.18 \\+07 04 33.5 } & 0.013 & 59$\pm$11 & 336, 342 & 8648, 8652 & \\
2005mf & UGC4798 & \linebreakcell[t]{09 08 42.33 \\+44 48 51.4 } & \linebreakcell[t]{5.9W\\13.3N} & Ic &&& 2005-12-25* & 17.3* & --- & 2005-12-31* & 18.00* & V* & \linebreakcell[t]{09 08 42.61 \\+44 48 38.3 } & 0.027 & --- & 335 & 8648, 8650 & \\
2005me & E244-31 & \linebreakcell[t]{01 30 09.22 \\-42 41 07.0 } & \linebreakcell[t]{41.E\\3.N} & II &&& 2005-12-23 & 17.5 & --- & --- & --- & --- & \linebreakcell[t]{01 30 05.43 \\-42 41 11.0 } & 0.022 & 82$\pm$16 & 333, 345, 2268 & 8647, 8652 & \\
2005mc & UGC4414 & \linebreakcell[t]{08 27 06.36 \\+21 38 45.6 } & \linebreakcell[t]{4.8E\\3.2N} & Ia &&& 2005-12-23 & 17.** & --- & --- & --- & --- & \linebreakcell[t]{08 27 05.97 \\+21 38 43.1 } & 0.025 & 115$\pm$9 & 331, 334, 338 & 8647 & \\
2005mb & NGC4963 & \linebreakcell[t]{13 05 52.46 \\+41 42 59.0 } & \linebreakcell[t]{5.5E\\20.2S} & II &&& 2005-12-21*** & 19.3 & --- & --- & --- & --- & \linebreakcell[t]{13 05 51.97 \\+41 43 19.1 } & 0.024 & --- & 330, 345 & 8647, 8652 & \\
2005ma & PGC16118 & \linebreakcell[t]{04 49 53.91 \\-10 45 23.4 } & \linebreakcell[t]{1.9W\\7.2S} & II* && n & 2005-12-24 & 17.9*** & --- & --- & --- & --- & \linebreakcell[t]{04 49 53.98 \\-10 45 16.1 } & 0.015 & 56$\pm$9 & 329, 334 & 8647 & \\
2005lz & UGC1666 & \linebreakcell[t]{02 10 49.76 \\+34 58 58.2 } & \linebreakcell[t]{13.8E\\6.3S} & Ia &&& 2005-12-24 & 17.8 & --- & --- & --- & --- & \linebreakcell[t]{02 10 48.61 \\+34 59 03.9 } & --- & 193$\pm$14 & 329, 337 & 8646 & \\
2005ly & UGC934 & \linebreakcell[t]{01 23 28.84 \\+30 46 45.7 } & \linebreakcell[t]{5.4E\\18.2S} & II && n & 2005-12-21 & 18.6*** & --- & --- & --- & --- & \linebreakcell[t]{01 23 28.34 \\+30 47 04.0 } & 0.035 & --- & 325, 327 & 8646 & \\
2005lx & IC221 & \linebreakcell[t]{02 22 39.01 \\+28 15 07.1 } & \linebreakcell[t]{24.8W\\17.9S} & II* &&& 2005-12-19 & 18.1*** & --- & --- & --- & --- & \linebreakcell[t]{02 22 40.90 \\+28 15 25.1 } & 0.017 & 67$\pm$11 & 322, 342, 1175 & 8646, 8652 & \\
2005lw & IC672 & \linebreakcell[t]{11 08 03.21 \\-12 29 09.4 } & \linebreakcell[t]{1.5E\\6.5S} & II &&& 2005-12-14 & 18.8** & --- & --- & --- & --- & \linebreakcell[t]{11 08 03.23 \\-12 29 03.0 } & 0.026 & --- & 318, 321 & 8646 & \\
2005lv* & UGC2964 & \linebreakcell[t]{04 08 56.47 \\+27 11 50.8 } & \linebreakcell[t]{13.5W\\2.4S} & --- &&& 2005-12-06 & 18.7 & --- & --- & --- & --- & \linebreakcell[t]{04 08 57.44 \\+27 11 52.1 } & 0.03 & --- & --- & 8645 & \\
2005lu & E545-38 & \linebreakcell[t]{02 36 03.71 \\-17 15 50.0 } & \linebreakcell[t]{1.8E\\2.2S} & Ia &&& 2005-12-11 & 17.1** & --- & --- & --- & --- & \linebreakcell[t]{02 36 03.60 \\-17 15 46.8 } & 0.032 & 152$\pm$2 & 317, 321 & 8645, 8652 & \\
2005lt & PGC36349 & \linebreakcell[t]{11 42 26.21 \\+20 07 05.0 } & \linebreakcell[t]{24.5E\\5.4S} & Ia &&& 2005-12-10 & 15.7 & --- & --- & --- & --- & \linebreakcell[t]{11 42 24.50 \\+20 07 09.4 } & 0.02 & 80$\pm$14 & 316, 319, 321 & 8645, 8652 & \\
2005ls & PGC10948 & \linebreakcell[t]{02 54 15.97 \\+42 43 29.8 } & \linebreakcell[t]{6.6W\\2.9S} & Ia &&& 2005-12-09 & 15.8* & --- & 2005-12-13* & 15.8* & --- & \linebreakcell[t]{02 54 16.52 \\+42 43 33.5 } & 0.021 & 74$\pm$7 & 324 & 8643, 8652 & \\
2005lr & E492-2 & \linebreakcell[t]{07 11 39.03 \\-26 42 20.2 } & \linebreakcell[t]{18.8W\\2.1S} & Ic* &&& 2005-12-04 & 18.5 & --- & --- & --- & --- & \linebreakcell[t]{07 11 40.40 \\-26 42 18.0 } & 0.0086 & --- & 321 & 8641, 8652 & \\
2005lq & --- & \linebreakcell[t]{02 41 36.04 \\+00 12 18.1 } & --- & Ia &&& 2005-11-24 & 22.7* & g* & 2005-12-01* & 22.3* & g* & \linebreakcell[t]{02 41 36 \\+00 12 18 } & 0.37**** & --- & 315 & 8640 & \\
2005lp & --- & \linebreakcell[t]{01 47 42.80 \\+00 12 26.0 } & --- & Ia &&& 2005-11-24 & 22.2* & g* & 2005-11-15* & 22.1* & g* & \linebreakcell[t]{01 47 42 \\+00 12 26 } & 0.3**** & --- & 315 & --- & \\

\end{longtable}
\end{landscape}

\end{appendix}

\end{document}